\documentclass[apjl]{emulateapj}
\usepackage{apjfonts}

\usepackage{apjfonts}

\newcommand{\lat}{\textit{Fermi}-LAT}
\newcommand{\pks}{PKS~0426$-$380}

\shorttitle{VHE $\gamma$ rays from \pks}
\shortauthors{Tanaka et al.}

\begin{document}

\title{{\it Fermi} Large Area Telescope Detection of Two Very-High-Energy ($E>100$\,GeV) $\gamma$-ray Photons from the $z = 1.1$ Blazar PKS~0426$-$380}

\author{Y.~T.~Tanaka\altaffilmark{1}, C.~C.~Cheung\altaffilmark{2}, Y.~Inoue\altaffilmark{3}, {\L}.~Stawarz\altaffilmark{4,5}, M.~Ajello\altaffilmark{6}, C.~D.~Dermer\altaffilmark{2}, D.~L.~Wood\altaffilmark{7}, A.~Chekhtman\altaffilmark{8}, Y.~Fukazawa\altaffilmark{9}, T.~Mizuno\altaffilmark{1}, M.~Ohno\altaffilmark{9}, D.~Paneque\altaffilmark{10,3}, D.~J.~Thompson\altaffilmark{11}}

\altaffiltext{1}{Hiroshima Astrophysical Science Center, Hiroshima University, Higashi-Hiroshima, Hiroshima 739-8526, Japan}
\altaffiltext{2}{Space Science Division, Naval Research Laboratory, Washington, DC 20375-5352, USA}
\altaffiltext{3}{W. W. Hansen Experimental Physics Laboratory, Kavli Institute for Particle Astrophysics and Cosmology, Department of Physics and SLAC National Accelerator Laboratory, Stanford University, Stanford, CA 94305, USA}
\altaffiltext{4}{Institute of Space and Astronautical Science, JAXA, 3-1-1 Yoshinodai, Chuo-ku, Sagamihara, Kanagawa 252-5210, Japan}
\altaffiltext{5}{Astronomical Observatory, Jagiellonian University, 30-244 Krak\'ow, Poland}
\altaffiltext{6}{Space Sciences Laboratory, 7 Gauss Way, University of California, Berkeley, CA 94720-7450, USA}
\altaffiltext{7}{Praxis Inc., Alexandria, VA 22303, resident at Naval Research Laboratory, Washington, DC 20375, USA}
\altaffiltext{8}{Center for Earth Observing and Space Research, College of Science, George Mason University, Fairfax, VA 22030, resident at Naval Research Laboratory, Washington, DC 20375, USA}
\altaffiltext{9}{Department of Physical Sciences, Hiroshima University, Higashi-Hiroshima, Hiroshima 739-8526, Japan}
\altaffiltext{10}{Max-Planck-Institut f\"ur Physik, D-80805 M\"unchen, Germany}
\altaffiltext{11}{NASA Goddard Space Flight Center, Greenbelt, MD 20771, USA}

\email{ytanaka@hep01.hepl.hiroshima-u.ac.jp}

\begin{abstract}
We report the {\it Fermi} Large Area Telescope (LAT) detection of two very-high-energy (VHE, $E>100$\,GeV) $\gamma$-ray photons from the directional vicinity of the distant (redshift, $z = 1.1$)  blazar \pks. The null hypothesis that both the 134 and 122 GeV photons originate from unrelated sources can be rejected at the $5.5 \sigma$ confidence level. We therefore claim that at least one of the two VHE photons is securely associated with \pks, making it the most distant VHE emitter known to date. The results are in agreement with recent \lat\ constraints on the Extragalactic Background Light (EBL) intensity, which imply a $z \simeq 1$ horizon for $\simeq 100$\,GeV photons. The LAT detection of the two VHE $\gamma$-rays coincided roughly with flaring states of the source, although we did not find an exact correspondence between the VHE photon arrival times and the flux maxima at lower $\gamma$-ray energies. Modeling the $\gamma$-ray continuum of \pks\ with daily bins revealed a significant spectral hardening around the time of the first VHE event detection (LAT photon index $\Gamma \simeq 1.4$) but on the other hand no pronounced spectral changes near the detection time of the second one. This combination implies a rather complex variability pattern of the source in $\gamma$ rays during the flaring epochs. An additional flat component is possibly present above several tens of GeV in the EBL-corrected \lat\ spectrum accumulated over the $\sim 8$-month high state.
\end{abstract}


\section{Introduction}
Blazars are radio-loud active galactic nuclei (AGN) with relativistic jets viewed at small angles to the Earth's line of sight. Their broad-band spectral energy distributions (SEDs) are typically dominated by two non-thermal emission components widely believed to be due to synchrotron and inverse-Compton emission from a single population of ultra-relativistic electrons accelerated within the innermost parts of the jets \citep[e.g.,][]{Urry95}. The exact particle acceleration processes at work, as well as the location and structure of the dominant energy dissipation zone in blazar sources (hereafter `the blazar zone'), are still under debate. Blazars are typically sub-divided into Flat Spectrum Radio Quasars (FSRQs) and BL Lacertae objects (BL Lacs) based on the equivalent widths of the emission lines in their optical spectra \citep[e.g.,][]{Urry95}. BL Lacs are characterized by much weaker emission lines than FSRQs, or even featureless optical continua, that can be understood in terms of distinct accretion rates in an otherwise homogeneous population of sources \citep[e.g.,][]{Ghisellini11}.

Since the launch of the {\it Fermi} satellite in 2008, high-redshift blazars have been established as very-high-energy (VHE; $> 100$\,GeV) emitters \citep{Neronov11, Neronov12}, with the highest-redshift case reported being a 126 GeV photon from B2 0912+29, a z = 1.521 BL Lac, though the redshift is still uncertain \citep[see][]{Neronov12}.
Although collectively detected, individually these associations are based on single photons and the chance coincidence probabilities for the VHE events to be detected around the sources are not low enough to claim discovery of a single source. Detection of sub-TeV VHE photons from $z \sim 1$ blazars is in principle reconcilable with most of the recently refined EBL models \citep{F08,Dominguez11,Stecker12,Yoshi13}.

Here, we report the discovery of VHE $\gamma$-ray emission from the blazar \pks, classified early on as a BL Lac \citep[e.g.][]{Sba05}, and only recently recognized as a FSRQ based on a new classification scheme advocated by \citet{Ghisellini11} and \citet{Sbarrato12}. These authors proposed that the dividing line between the two types of blazars corresponds to the critical BLR luminosity in the Eddington units, $L_{\rm BLR}/L_{\rm Edd} \simeq 5 \times 10^{-4}$. In the case of \pks, the precise characterization of the optical spectrum by Very Large Telescope (VLT) observations revealed a broad [Mg\,II] line with an equivalent width of 5.7\,\AA\ and a relatively high luminosity of $7.2 \times 10^{42}$\,erg\,s$^{-1}$ \citep{Sba05}. This leads to $L_{\rm BLR}/L_{\rm Edd} \simeq 10^{-3}$ for a black hole mass $\mathcal{M}_{\rm BH} \simeq 10^{9}\,M_{\odot}$ (see Section 4), and therefore the FSRQ classification according to the proposed scheme \citep{Ghisellini11}.

The VLT detection of the broad [Mg\,II] $\lambda$2798 emission line mentioned above by \citet{Sba05}, together with C\,III] and  [O\,II] $\lambda$3727, enabled the determination of a redshift, $z=1.105$, for the source ($d_L \simeq 7.52$\,Gpc, assuming standard cosmology with $H_0=71$\,km\,s$^{-1}$\,Mpc$^{-1}$, $\Omega_m=0.27$, and $\Omega_{\Lambda}=0.73$). Independently, \citet{Heidt04} also derived $z=1.111$ based on the single [Mg\,II] $\lambda$2798 emission line; the non-detection of a host galaxy in the {\it Hubble} Space Telescope image \citep{Urry00} is consistent with the high redshift. Given all these findings, we conclude that the redshift determination for \pks\ is robust.

The \lat\ detection of a VHE event from near the direction of \pks\ in January 2010 was previously reported \citep{2LAC,Neronov12}. However, the probability that the VHE event originated from other sources (including background/foreground diffuse emissions) was relatively large, and the significance did not reach $5 \sigma$. In this Letter, we report the \lat\ detection of a second VHE event from the directional vicinity of \pks\ in January 2013, and claim convincingly that it is now the most distant VHE emitter currently detected.

\begin{table*}
\noindent
{\caption[] {Detailed description of the two VHE events detected by \lat}}
\begin{center}
\begin{tabular}{cccccc}
\hline \hline
\\
Energy$^{\ast}$ & MET & R.A. (J2000) & Dec. (J2000) & Angular separation$^{\dagger}$ & {\tt gtsrcprob}$^{\ddagger}$ \\
 $[$GeV] & (UT) & [deg] & [deg] & [deg] & probability \\
\\
\hline
\\
134 & 285043901.724          & 67.182 & $-$37.930 & 0.013 & 0.9999763 \\
                & (2010 Jan 13 02:51:39.724) &                 &                       &               & \\
122 & 380539944.325          & 67.194 & $-$37.943 & 0.021 & 0.9999720 \\
                & (2013 Jan 22 09:32:21.325) &                 &                       &               & \\
\\
\hline\hline
\end{tabular}
\end{center}
\tablecomments{Both of the events are {\tt ULTRACLEAN} class and {\tt FRONT} converting.}
$^{\ast}$ The energy resolution is of the order of 10\% \citep{LATpass7}.

$^{\dagger}$ Angular separation is calculated from the radio position of \pks, R.A.=67.1684342$^{\circ}$ and Dec.=$-$37.9387719$^{\circ}$ (J2000) \citep{Johnston95}.

$^{\ddagger}$ The probability that the event belongs to \pks, which is calculated by using {\tt gtsrcprob}.
\end{table*}

\section{Data Reduction}
The \lat\ {\tt Pass7} event and spacecraft data ({\tt ft1} and {\tt ft2} files, respectively) were downloaded from the LAT Data Server\footnote{http://fermi.gsfc.nasa.gov/cgi-bin/ssc/LAT/LATDataQuery.cgi} at the {\it Fermi} Science Support Center webpage. We took the {\tt ft1} photon event data file spanning  Mission Elapsed Time (MET, measured in seconds from 2001 Jan. 1)   239557417 (2008 August 4 15:43:36 UT) to 389039485 (2013 April 30 18:31:23 UT) and chose a 10$^{\circ}$  radius for the region of interest (ROI) centered at the radio position of \pks. The event selection and data analysis were performed in a standard manner using version {\tt v9r27p1} of the {\it Fermi Science Tools}. Only the {\tt CLEAN} class events from 100\,MeV to 300\,GeV were selected. The maximum zenith angle was set to 100$^{\circ}$ to avoid contamination from Earth limb $\gamma$ rays. The good time interval was generated by applying a recommended filter expression of {\tt (DATA\_QUAL==1)\&\&(LAT\_CONFIG==1)\&\&ABS(ROCK\_} {\tt ANGLE)$<$52} and ROI-based zenith angle cut ({\tt roicut=yes}). We used the {\tt P7CLEAN\_V6} Instrument Response Functions (IRFs). 

We first performed unbinned maximum likelihood ({\tt gtlike}) analysis for the 4.7-year LAT data by using an XML source model\footnote{Models are defined using the XML language.} in which the spectral parameters of all the sources included in the Second \lat\ Catalog \citep[2FGL;][]{Nolan12} within 10$^{\circ}$ radius were set free, while those within an annulus from 10$^{\circ}$ to 15$^{\circ}$ were fixed to their 2FGL values. For \pks\ we assumed a log-parabola spectral shape, as measured in the 2FGL catalog. The template files {\tt gal\_2yearp7v6\_v0.fits} and {\tt iso\_p7v6clean.txt} were used to represent the Galactic and isotropic diffuse emission components, respectively\footnote{Available from the {\it Fermi} Science Support Center (FSSC), http://fermi.gsfc.nasa.gov/ssc/data/access/lat/BackgroundModels.html}. To allow for potential small errors in the flux and spectrum of the Galactic diffuse emission model we multiplied it by a power law in energy whose normalization and index were free during the fit. Using the output XML file obtained after running {\tt gtlike}, we ran the {\tt gtsrcprob} tool to calculate a probability that each detected VHE event originates from the direction of \pks. For the same output XML file, we freed only the normalizations of the surrounding 2FGL sources and the Galactic and extragalactic diffuse emission components and modeled \pks\ with a single power-law with both the normalization and the photon index allowed to vary. Then, we ran {\tt gtlike} and generated a weekly (7-day binned) light curve for \pks. Finally, when calculating daily fluxes and power-law indices for \pks, only the source normalization and power-law slope were left free, along with the normalizations of the two diffuse emission components, while all the other parameters were fixed to the values obtained with {\tt gtlike} for the entire 4.7-year dataset to avoid unreasonably large errors.

To construct the $\gamma$-ray spectrum of the source, we first selected energy intervals chosen as 11 octaves ranging from 0.1 to 204.8\,GeV (namely, 0.1--0.2, 0.2--0.4, ..., 102.4--204.8\,GeV). In the XML source model, the normalizations of the surrounding sources within 10$^{\circ}$ and of the Galactic and extragalactic diffuse emission components were set free, while all the other spectral parameters were fixed to the values derived for the LAT data accumulated for the selected period. \pks\ was modeled using a broken power-law model with the photon indices and break energy fixed to the values that we derived for the entire dataset, but with the normalization set free. Using this XML file, we ran {\tt gtlike} for each energy bin and generated the spectrum of the source.

\section{Results}
In Figure 1 we present the LAT count map of {\tt ULTRACLEAN} events (a subset of the CLEAN class with the highest probability of being $\gamma$ rays) with 5--300\,GeV energies around \pks. One can clearly see that the two VHE photons, with energies 134\,GeV and 122\,GeV, coincide with the position of the blazar (angular separations from the blazar of $\leq 0.021^{\circ}$) and are well inside the point spread function (PSF) of \lat. A detailed summary of the VHE events, including their precise localizations and arrival times, is given in Table~1. Based on the {\tt gtsrcprob} results (see Table~1), we calculated the null hypothesis probability that both events originate from foreground/background (Galactic/extragalactic) diffuse emission components, or other surrounding 2FGL sources rather than \pks, using Fisher's method and obtained $1.47 \times 10^{-8}$. We can therefore reject the null hypothesis at the $5.5 \sigma$ confidence level (CL), and claim robustly that at least one of the two detected VHE $\gamma$ rays originates from \pks, making it the most distant VHE emitter known to date.

\begin{figure}
\begin{center}
\plotone{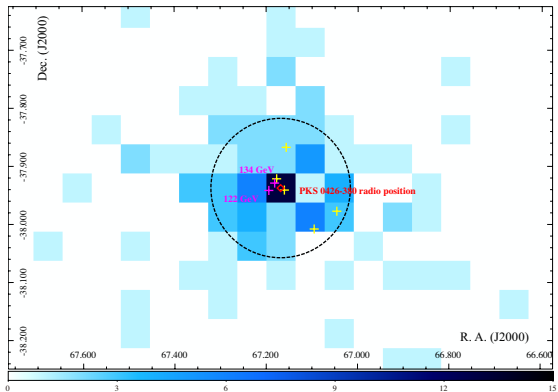}
\caption{\lat\ count map of 5--300\,GeV {\tt ULTRACLEAN} events (scale = 0.05$^{\circ}$ pixel$^{-1}$) centered on the radio position of \pks\ (red diamond). The magenta and yellow crosses indicate the positions of the two VHE and five 50--100\,GeV events, respectively. The dashed circle indicates the 68\% containment radius (0.12 deg) of \lat\ PSF for front-converting events above 100 GeV \citep{LATpass7}.}
\end{center}
\end{figure}

Although the two VHE events are classified as {\tt ULTRACLEAN} ones, we further visually inspected the tracks in the LAT tracker subsystem after the point of pair conversion. The first event converted at the top layer of a tower and the long straight paths of the converted electron-positron pairs were nicely tracked. The second event converted at the third layer of the tracker and similarly showed no unusual signatures. The showers in the calorimeters for both events were also well-behaved. In conclusion, we did not find any problematic features in the case of the analyzed VHE detections.

Figure~2 presents the weekly (7-day binned) \lat\ light curve of \pks\ within the energy range $0.1-300$\,GeV, along with the arrival times of $\gamma$ rays with energies above 50 GeV. It is clear that the two VHE events were detected during high states of the source. Interestingly however, the VHE detections did not coincide with the flux maxima of the flaring epochs. Complex structures of the source light curve, with multiple peaks within each flaring state, preclude us however from speculating if the VHE photons preceded or followed the flux maxima at lower photon energies. To further investigate the temporal and spectral variations within $\pm 10$ days of the two VHE events, in Figure~3 we show the daily changes of the flux and power-law index for the source. In neither case did the $0.1-300$\,GeV flux dramatically increase on the day of the VHE detection. On the other hand, a significant spectral hardening can be noticed on the day of the first VHE detection, but not on the day of the second one.

\begin{figure}
\begin{center}
\plotone{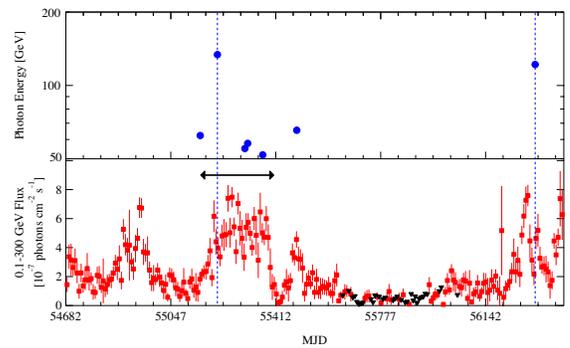}
\caption{{\it top} panel: Arrival times (MJD) and energies of the seven most energetic $E>50$\,GeV {\tt ULTRACLEAN} events detected from the vicinity of the direction of \pks\ (see Figure~1). {\it bottom} panel: \lat\ weekly binned light curve of \pks\ ($0.1-300$\,GeV energy range). Black triangles denote the 95\% CL flux upper limits (when TS $< 10$). The two vertical dashed lines denote the VHE detection times (MJD 55209.11920977 and 56314.39746904; cf., Table~1). Also shown by a black horizontal line is the accumulation period for which the \lat\ spectrum was constructed (see Figure~4).}
\end{center}
\end{figure}

\begin{figure}
\begin{center}
\plotone{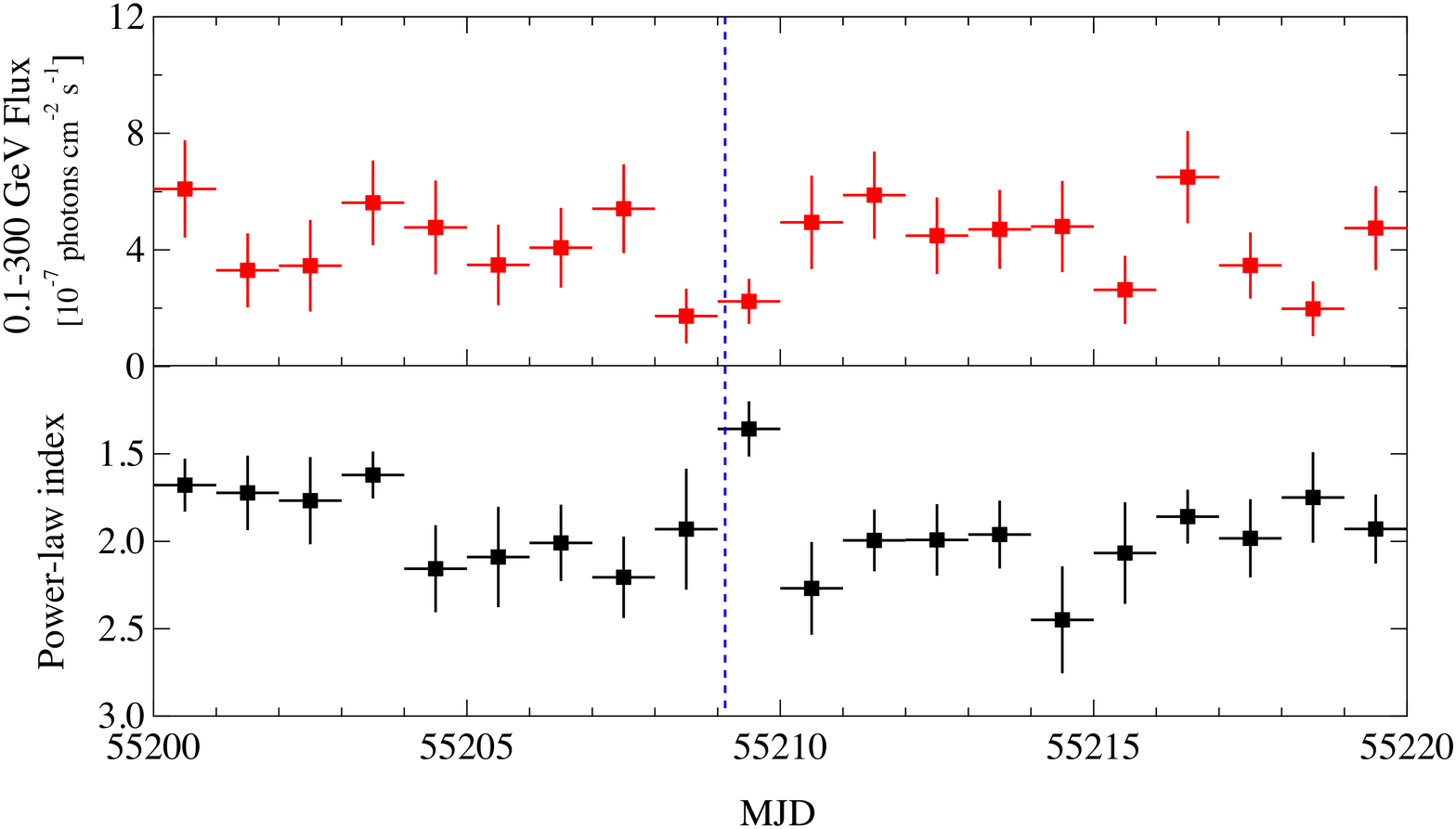}
\plotone{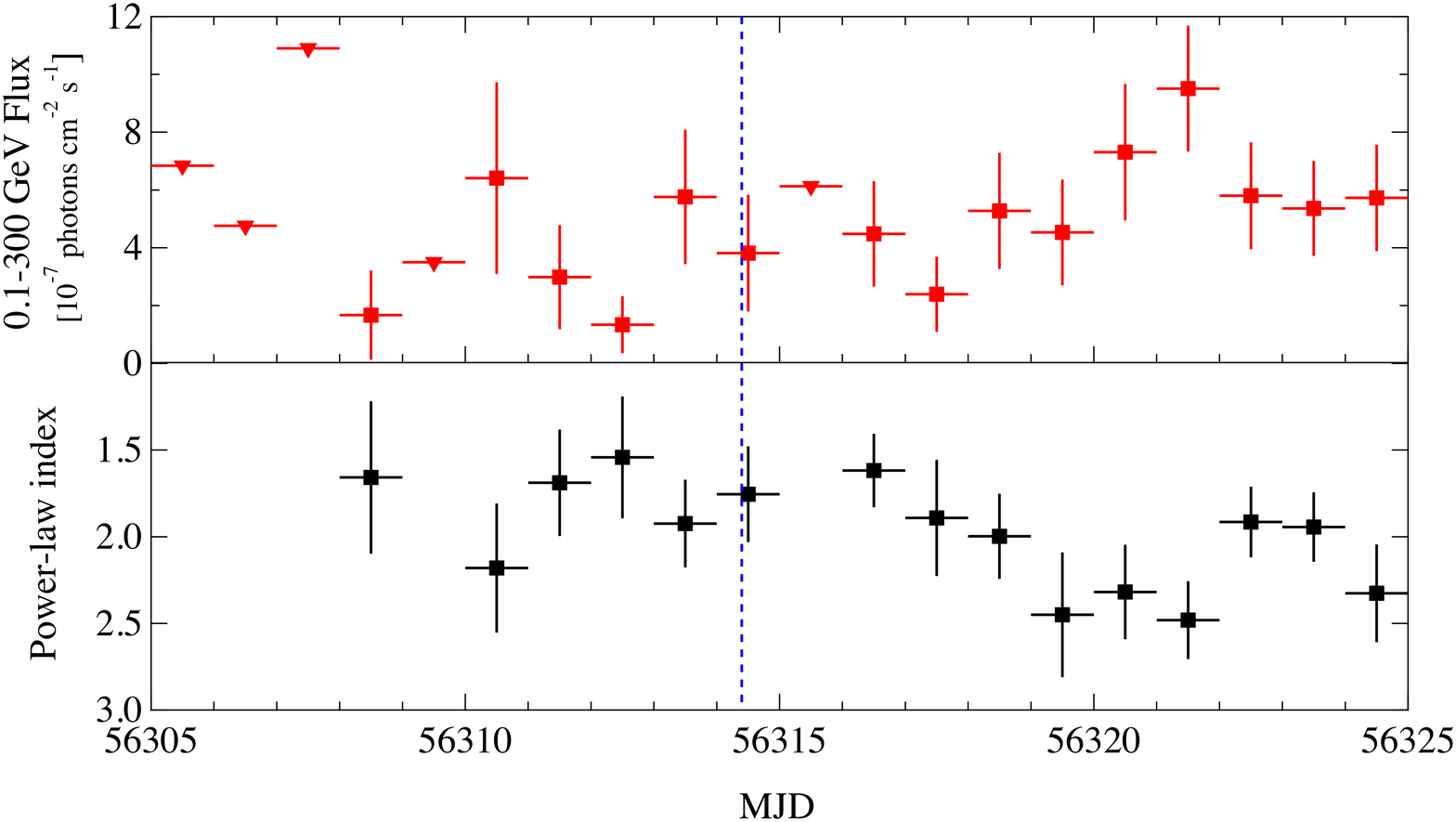}
\caption{\lat\ daily binned light curve and power-law indices (0.1--300\,GeV energy range) derived for \pks\ around the times ($\pm 10$ days) of the VHE detections as denoted in the figure by blue vertical dashed lines (MJD 55209.11920977 and 56314.39746904). Triangles without vertical error bars denote the 95\% CL flux upper limits when TS$<10$. Note the 4.7-year average 0.1--300\,GeV photon flux was $\left( 1.8 \pm 0.1 \right) \times 10^{-7}$ photons cm$^{-2}$ s$^{-1}$.}
\end{center}
\end{figure}

Figure~4 shows the $\gamma$-ray spectrum of \pks\ during the most energetic flare, derived from the accumulated \lat\ data between MET 280000000 and 302000000 (see Figure~2). No high-energy cutoff expected from the EBL-related attenuation of the $\gamma$-ray continuum seems to be present up to several tens of GeV, although the highest energy ($102.4-204.8$\,GeV) datapoint is a 95\% CL upper limit due to the limited photon statistics (corresponding to a single net photon). We found that a broken power-law model of $\Gamma_{\rm low}=1.91\pm0.04$, $\Gamma_{\rm high}=2.72\pm0.17$, and $E_{\rm break}=8.0\pm0.9$\,GeV maximized the likelihood, and this is also drawn in Figure~4. We emphasize that the presented spectrum was derived from the \lat\ data accumulated over $\sim 8$ months. Hence, keeping in mind the large-amplitude variability of the source in $\gamma$ rays, one has to be very careful in interpreting spectral features apparent in the figure, like for example the discontinuity around 10\,GeV \citep[see][for other examples]{Abdo10}. Based on the observed SED, we generated the EBL-corrected spectra of the source (Figure~4) using two different EBL models \citep{F08,Yoshi13}. The de-absorbed spectra seem to reveal an additional high-energy flat-spectrum component above several tens of GeV \citep[see also][]{Senturk13}. However, the significance of this feature is marginal and, more importantly, model dependent. Nonetheless, the results obtained do suggest that \pks\ is a very promising target for future follow-up studies with Imaging Atmospheric Cherenkov Telescopes (IACTs) such as H.E.S.S. II and CTA \citep{CTA}.

\section{Discussion}
Three blazars classified as FSRQs have previously been detected in the VHE range by IACTs, 3C~279 \citep[$z=0.536$;][]{279MAGIC}, PKS~1510$-$089 \citep[$z=0.361$;][]{1510HESS}, and 4C~+21.35 \citep[$z=0.432$;][]{Aleksic11}. VHE emissions from more distant objects such as KUV~00311$-$1938 at $z=0.61$ \citep[though the redshift is still tentative]{Becherini12} and PKS~1424+240 at $z \geq 0.6035$ \citep{Furniss13} have been recently detected. The observational results presented in this Letter therefore establish \pks, which is located at $z=1.1$, as the most distant VHE emitter observed to date. We note that the redshift of \pks\ is just around the horizon for $\simeq 100$\,GeV $\gamma$ rays (namely, EBL-related optical depth of the Universe $\tau_{100\,{\rm GeV}} \sim 1$) as recently determined by \lat\ \citep[and Figure 1 therein]{Marco12}, hence the VHE detection from the $z=1.1$ blazar is not unreasonable.

\begin{figure}[t]
\begin{center}
\plotone{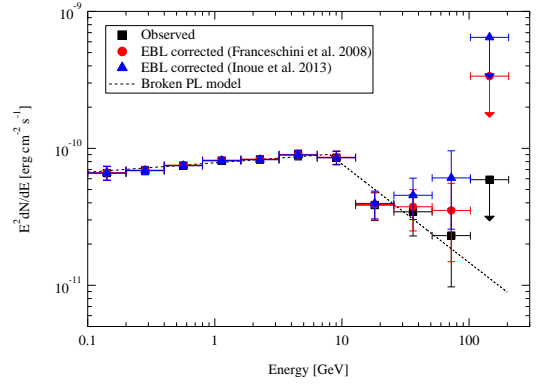}
\caption{SED of \pks\ derived from the \lat\ data accumulated during the most energetic flaring state spanning  MET 280000000 (17:46:38 UT on 2009 November 15) to 302000000 (08:53:18 UT on 2010 July 28; see also the black horizontal line in Figure~2). The observed spectrum is denoted by black squares and the highest energy bin is a 95\% confidence level upper limit. 
A broken power-law model which maximizes the likelihood for the 0.1--300 GeV \lat\ data is indicated with black dashed line. The spectra corrected for the EBL-related attenuation, using the EBL models of \citet{F08} and \citet{Yoshi13}, are represented by red circles and blue triangles, respectively.}
\end{center}
\end{figure}

As mentioned previously, the observed luminosity of the broad [Mg\,II] line in the optical spectrum of \pks\ is $L_{\rm Mg\ II} \simeq 7.2 \times 10^{42}$\,erg\,s$^{-1}$ \citep{Sba05}, implying a BLR luminosity $L_{\rm BLR} \simeq 1.2 \times 10^{44}$\,erg\,s$^{-1}$ using the scaling relation $L_{\rm BLR} \simeq 16.4 \times L_{\rm Mg\ II}$ \citep{Wang04}. Spectral modeling of the accretion-related continuum in the source \citep{Ghisellini11,Sbarrato12} yields $\mathcal{M}_{\rm BH} \simeq 4 \times 10^{8} M_{\odot}$. Based on the [Mg\,II] line FWHM of $4,700$\,km\,s$^{-1}$, and the observed $V$-band magnitude of 18.6 (assuming negligible starlight contamination), we derived a slightly larger value of $\mathcal{M}_{\rm BH} \simeq (0.9-1.3) \times 10^{9} M_{\odot}$, using the scaling relations from \citet{Wang09} and \citet{Vest09}. This corresponds to the Eddington luminosity $L_{\rm Edd} \simeq 10^{47}$\,erg\,s$^{-1}$, the ratio $L_{\rm BLR}/L_{\rm Edd} \simeq 10^{-3}$, and the accretion rate at the level of $\Lambda_{\rm acc} = L_{\rm disk} / L_{\rm Edd} \sim 30\%$ (assuming the standard bolometric correction factor $L_{\rm disk} \simeq 10 \times L_B$ for the $B$-band source luminosity $L_B \simeq 4.5 \times 10^{45}$\,erg\,s$^{-1}$). The derived high accretion rate is consistent with \pks\ being a FSRQ.

The intense $\gamma$-ray emission of FSRQs is widely thought to arise due to inverse-Compton up-scattering of low-energy photons generated outside of a jet by ultra-relativistic electrons accelerated within the innermost parts of the relativistic outflows \citep{Sikora09,Ghisellini09}. If this blazar emission zone is located at sub-parsec distances from the central black hole, as is often anticipated in the literature, then the abundant circum-nuclear photon fields provided by the BLR and/or hot dust are expected to attenuate the VHE blazar emission substantially due to the photon-photon pair production, leading to the formation of breaks and cut-offs in the $\gamma$-ray continua of FSRQs \citep[see in this context the discussion in][]{Poutanen10,Tanaka11}. The observed sharp break ($\Gamma_{\rm high} - \Gamma_{\rm low} \sim 0.8$) at $\sim 8$\,GeV would be understood by a scenario with the blazar emission zone deep within the BLR. However, we note again that the $\gamma$-ray spectrum when corrected for the cosmological absorption showed a flattened shape at energies above 10 GeV. This flat component, if connected to the sub-TeV range, should come from another emission region outside the BLR to avoid $\gamma\gamma$ attenuation.

Care must be taken not to over-interpret such high-energy features (flattening) in unfolded blazar spectra constructed using \lat\ data accumulated over longer periods of time, since those may simply arise due to averaging over different activity states characterized by different spectral properties \citep[see the related discussion and analysis in][concerning the well-known BL Lac object Mrk~501]{Abdo2011mrk501}. Still, the results presented here for \pks\ are in principle consistent with the emergence of an additional very high-energy flat-spectrum component during the flaring states of the source. One possibility for the production of such a component could be electron pile-up at the highest energies due to the efficient and continuous acceleration processes limited only by the radiative losses \citep{Stawarz08,Lefa}. Another possibility could be an additional hadronic emission component dominating occasionally the source spectrum in the VHE range \citep[e.g.,][]{Boettcher,Dermer12}.  In particular, assuming that \pks\ generates cosmic rays with energies above 10\,EeV, or $\gamma$ rays above 30\,TeV, the induced intergalactic cascade emission may provide a non-negligible contribution within the VHE range of the source spectrum \citep{ess10,mur12,takami13}. The time scale of the expected delay between the production of the primary ultra-high energy cosmic rays/$\gamma$-ray photons and the observed re-processed VHE signal is $\sim 3 \, (E/100\,{\rm GeV})^{-2} \, (B/10^{-18}\,{\rm G})^2 \, (\lambda_{\gamma\gamma}/100 \, {\rm Mpc})$\,days and $\sim 30\,(E/100\ {\rm GeV})^{-2} \, (B/10^{-18}\,{\rm G})^2 \, (\lambda_{\rm BH}/1 \, {\rm Gpc})$\,days for the photon and the cosmic-ray induced cascade emission at $z=1.1$, respectively, where $B$ is the intergalactic magnetic field strength, $\lambda_{\gamma\gamma}$ is the mean free path of the pair creation, and $\lambda_{\rm BH}$ is the mean free path of the Bethe-Heitler process \citep{murase12}. Rather speculatively, if there are axion-like particles, photon and axion mixing in the intergalactic medium may in addition enhance the EBL-absorbed photon flux \citep[e.g.,][]{san09}. Lorentz invariance violation, on the other hand, could inhibit pair production \citep[e.g.,][]{pro00}. 

In summary, with \lat\ we have detected two VHE $\gamma$ rays from close to the direction of a high-redshift FSRQ, \pks\ at $z=1.1$. Both of the events were detected during high $\gamma$-ray states of the source, although the specific VHE arrival times did not coincide with any particularly large $0.1-300$\,GeV source flux. Only in the first case was the observed LAT spectrum harder than usual. The very complex $\gamma$-ray variability patterns revealed by the \lat\ data for \pks\ calls for sensitive follow-up studies with simultaneous VHE coverage provided by IACTs such as H.E.S.S. II or future CTA.

\acknowledgments We appreciate the referee's critical reading and valuable comments. Y.T.T. is supported by Kakenhi 24840031. Work by C.C.C. at NRL is supported in part by NASA DPR S-15633-Y. \L .S. was supported by Polish NSC grant DEC-2012/04/A/ST9/00083.

The \lat Collaboration acknowledges support from a number of agencies and institutes for both development and the operation of the LAT as well as scientific data analysis. These include NASA and DOE in the United States, CEA/Irfu and IN2P3/CNRS in France, ASI and INFN in Italy, MEXT, KEK, and JAXA in Japan, and the K.~A.~Wallenberg Foundation, the Swedish Research Council and the National Space Board in Sweden. Additional support from INAF in Italy and CNES in France for science analysis during the operations phase is also gratefully acknowledged.

{\it Facilities:} \facility{Fermi (LAT)}.

\bibliographystyle{apj}

\end{document}